\documentclass[twocolumn,showpacs,showkeys, preprintnumbers,amsmath,amssymb]{revtex4-1} 

	\usepackage[utf8]{inputenc}
	\usepackage[T1]{fontenc}
	\PassOptionsToPackage{hyphens}{url} % Correct line breaks of URLs
	\usepackage[colorlinks=true, pdfborder={0 0 0}]{hyperref}
	\usepackage{url}
	\makeatletter\newcommand\multiabel[1]{\quad \refstepcounter{equation}(\theequation)\ltx@label{#1}}\makeatother
\usepackage{graphicx}
\usepackage{xcolor}
\usepackage{enumitem}
\usepackage{empheq}
\usepackage{bbold}

\pdfpageattr{/Group << /S /Transparency /I true /CS /DeviceRGB>>} % include for transparency issues of figures

\makeatletter
\renewcommand{\p@subsection}{}
\renewcommand{\p@subsubsection}{}
\makeatother

\begin{document}

\title{Disentangling Entanglements in Biopolymer Solutions}

\author{Philipp Lang}
\author{Erwin Frey}

\email[]{frey@lmu.de}
%\date{\today}

\affiliation{
$^{1}$Arnold Sommerfeld Center for Theoretical Physics and Center for NanoScience, Department of Physics, Ludwig-Maximilians-Universit\"at M\"unchen, Theresienstrasse 37, 80333 M\"unchen, Germany}

\begin{abstract}
Reptation theory has been highly successful in explaining the unusual material properties of entangled polymer solutions. 
It reduces the complex many-body dynamics to a single-polymer description where each polymer is envisaged to be confined to a tube through which it moves in a snake-like fashion. 
For flexible polymers, reptation theory has been amply confirmed by both experiments and simulations. 
In contrast, for semiflexible polymers experimental and numerical tests are either limited to the onset of reptation, or were performed for tracer polymers in a fixed, static matrix. 
Here we report Brownian dynamics simulations of entangled solutions of semiflexible polymers, which show that curvilinear motion along a tube (reptation) is no longer the dominant mode of dynamics.
Instead, we find that polymers disentangle due to correlated constraint release which leads to equilibration of internal bending modes before polymers diffuse the full tube length. The physical mechanism underlying terminal stress relaxation is rotational diffusion mediated by disentanglement rather than curvilinear motion along a tube. 
\end{abstract}

\maketitle

Dense solutions of polymers are viscoelastic: 
While they respond like a fluid to low-frequency stresses, they act like a cross-linked elastic network at high frequencies. 
These intriguing material properties are attributed to the extended structure of polymers, which makes topological interactions particularly important:
Polymers can effortlessly slide past each other but are not allowed to cross each other's path. 
These `entanglements' mutually restrict the accessible configuration space, turning the dynamics of polymer solutions into a difficult many-body problem. 
Efforts to incorporate these features into a single-polymer mean-field model led to the famous tube model\cite{edwards:67, degennes:71, doi:78}.
In this model, the dynamic topological constraints on the motion of a given polymer are represented as a static, confining tube\cite{edwards:67}.
By this means, the single-polymer dynamics in an entangled solution is reduced to the curvilinear Brownian motion of its center-of-mass along the long axis of the tube\cite{degennes:71}, termed `reptation'.
For flexible polymers, reptation theory is well established and also fairly predictive\cite{kremer1990dynamics, mcleish:02, everaers:04, hou2010stress}, though there are still many interesting open questions\cite{rubinstein:10}.

Reptation theory has also been employed to elucidate the viscoelastic properties of entangled solutions of semiflexible polymers, which play an important role in determining the material properties of biopolymer solutions\cite{broedersz:14}. 
While scaling theories for the tube width\cite{odijk:83, semenov:86} have been convincingly verified, both experimentally\cite{romanowska:09} and numerically\cite{ramanathan:07}, the predictions of reptation theory for the long-time dynamics and the ensuing (terminal) stress relaxation remain controversial for several reasons. 
First, different scaling theories lead to conflicting results for the dependence of the terminal relaxation time $\tau_r$ on polymer length $L$, persistence length $\ell_p$, and mesh size $\xi$: 
Results range from\cite{doi:78}  $\tau_{r} \,{\sim}\,  L^7 / \xi^4$ through\cite{odijk:83} $\tau_{r}\,{\sim}\, \ell_p L^2$  to\cite{granek1997semi} $\tau_{r} \,{\sim}\,  (\ell_p/\xi)^{2/3} L^3$. 
Recent experimental studies\cite{fakhri:10} of the Brownian motion of carbon nanotubes in porous agarose networks seem to support Odijk's scaling result\cite{doi:78} $\tau_r \,{\sim}\, \ell_p L^2$. 
However, these experimental results do not settle the actual controversy, as the polymer diffuses in a fixed, static matrix and not in an entangled polymer solution. 
Experiments on entangled solutions are sparse and mostly investigate the dynamics in the plateau regime, where polymers experience tube confinement but do not yet show curvilinear motion along the tube\cite{liu2006microrheology, semmrich:07, PerkinsChu1994Science}. 
Unfortunately, these studies provide no explicit information on the dependence of the terminal relaxation time on polymer length and persistence length\cite{hinner:98}. 
Furthermore, whether or not reptation, i.e.\ curvilinear motion along some `primitive path', is the actual mechanism of stress relaxation has never been tested experimentally or by means of computer simulations. 
While there are experiments reporting the direct observation of filament dynamics within a tube and sliding motion along the tube\cite{kaes:94, perkins:94}, following the polymer dynamics over longer time scales poses a formidable experimental challenge and has not yet been realized. 
Similarly, previous Brownian dynamics simulations span the regime up to intermediate time scales where the tube forms, but do not extend deeply enough into the terminal relaxation regime\cite{ramanathan:07}.

\begin{figure}[h!]
\centering
\includegraphics[width=\linewidth]{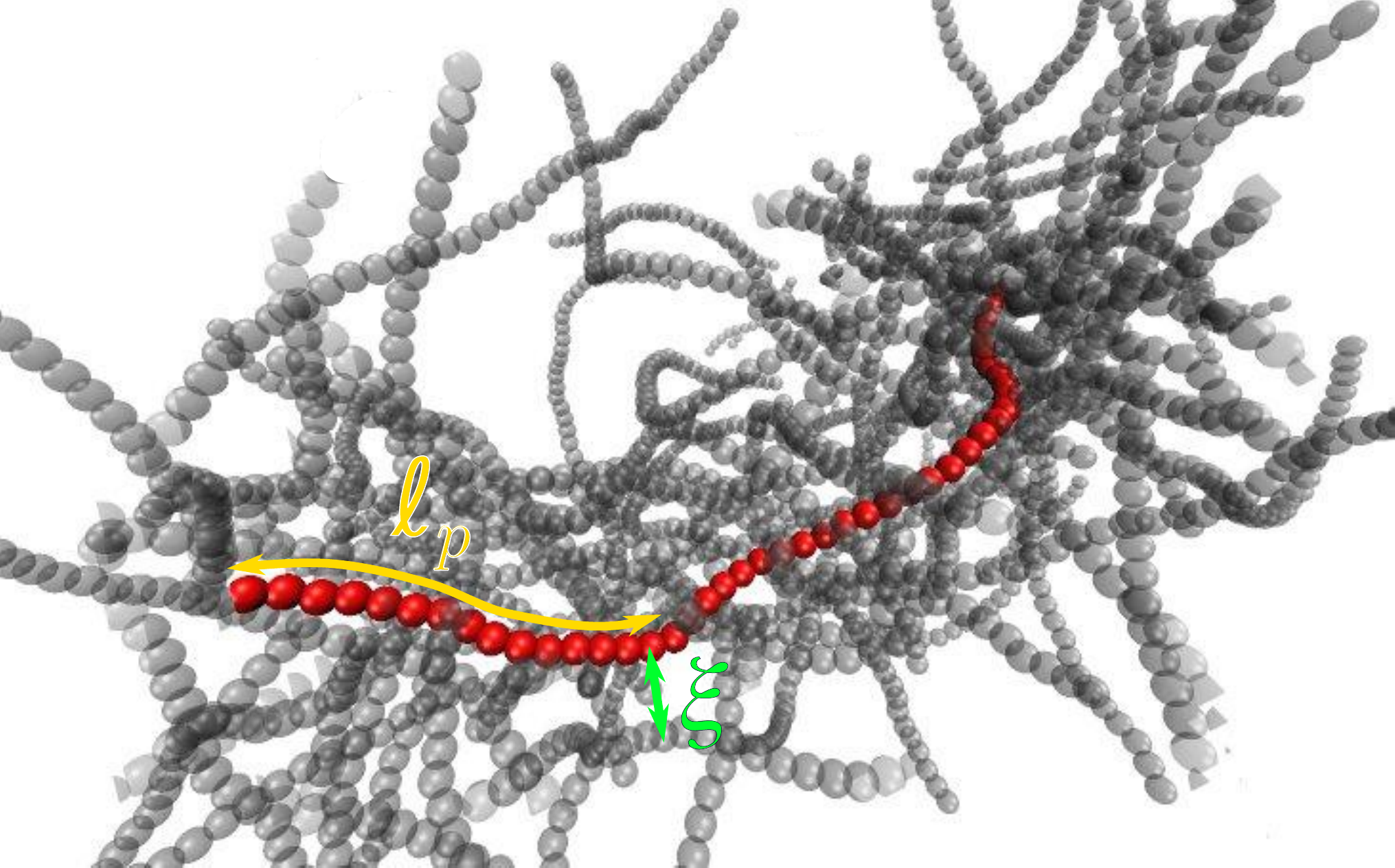}
\caption{
\textbf{Illustration of an entangled polymer solution.} Typical configuration of a tracer filament (red) with persistence length $\ell_p /L \,{=}\, 0.44$ in an entangled solution with mesh size $\xi/L \,{=}\, 0.1$. For better visibility of the surrounding chains (grey) only those in direct proximity to the tracer filament are shown. The dynamics of an entangled solution on different time scales is illustrated in the Supplementary Movie 1.
\label{fig:parameters}}
\end{figure} 

Here we present a large-scale Brownian dynamics simulation study and a complementary scaling approach for the dynamics of entangled solutions of semiflexible polymers. 
We consider a monodisperse solution of worm-like chains with length $L$ and persistence length $\ell_p$ that interact by volume exclusion only. 
The polymer density is given by the number $\nu$ of polymers per unit volume, with the mesh size defined as $\xi \,{:=}\, \sqrt{\frac{3}{\nu L}}$; for an illustration refer to Fig.~\ref{fig:parameters}. 
We are mainly interested in the dynamics of semiflexible polymers ($\ell_p \,{\gtrsim}\,  L$) at densities where the polymer length is much larger than the mesh size, $L \,{\gg}\, \xi$, but lies below the threshold of the isotropic-to-nematic transition. 
For example, in a typical solution of actin filaments of length $L  \,{\approx}\,  10 \, \mu$m at a number density
%$c  \,{\approx}\,  0.5$~mg/ml; \cite{schmidt:89}
$\nu  \,{\approx}\,  1 \, \mu$m$^{-3}$  the average mesh size is $\xi  \,{\approx}\,  0.5 \, \mu$m, and therefore much smaller than both $L$ and $\ell_p  \,{\approx}\,  17 \, \mu$m\cite{leGoffFrey2002prl}. 
The polymer dynamics is governed by a Langevin equation, 

\begin{equation}
	\zeta \partial_t \mathbf{r}_{i}^{k}  
	=  {-} \frac{\partial U}{\partial \mathbf{r}_{i}^{k}} {+} \boldsymbol{\eta} \, ,
\end{equation}
where $\mathbf{r}_{i}^{k}(t)$ is the position vector of bead $i \,{\in}\, \{ 1, \ldots, N \}$ on polymer $k$. 
The potential $U$ accounts for the bending energy of each filament as well as the mutual steric interaction between the filaments, $\zeta$ denotes the friction coefficient, and $\boldsymbol{\eta}$ is Gaussian white noise with an amplitude determined by the Stokes-Einstein relation; for further details see Methods.
Our Brownian dynamics simulations employ a standard bead-spring algorithm, as explained in Methods. 

\section*{Results}

To learn how the polymers first entangle and then disentangle, i.e. how topological constraints emerge and are released, we studied the dynamics of individual (tracer) polymers in  an entangled polymer solution; see Supplementary Movie 1.
We began our analysis by measuring the time evolution of the mean-squared displacement (MSD) of the end-to-end distance (Methods), $\delta R^2 (t)$, thus focusing  on the relaxation of the \textit{internal modes} and disregarding any global translation or rotation of the polymer.
\begin{figure}[t]
\centering
\includegraphics[width=\linewidth]{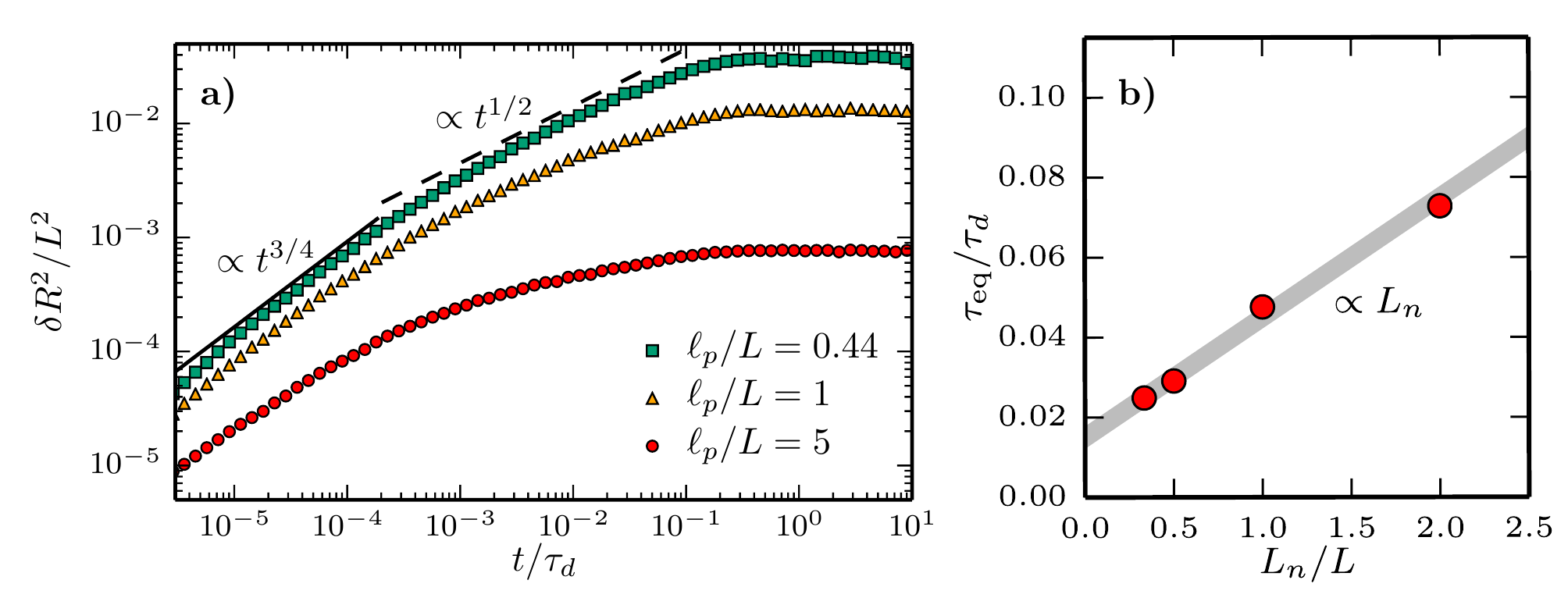}
\caption{
\textbf{Dynamics of the mean-square end-to-end distance.} {\bf a}, $\delta R^2 (t)$ of a single filament in an entangled polymer solution, with time rescaled in units of the diffusion time $\tau_d  \,{=}\,  \zeta L^3 / (6 k_B T)$ where $\zeta$ is the friction coefficient and $T$ the temperature. Simulations were performed for filaments with the indicated persistence lengths in a solution with mesh size $\xi/L \,{=}\, 0.086$. In each case, there are three distinct dynamic regimes: free relaxation following a $t^{3/4}$ power law, an intermediate regime where the tube reorganises and topological constraints are relaxed, and a final regime where bending modes are fully relaxed at their respective equilibrium value $\delta R_{eq}^2 (t) \,{=}\, L^4/(45 \ell_p^2)$. {\bf b}, Equilibration time of internal modes, $\tau_{eq}/\tau_d$, for a tracer filament of length $L$ (with $\xi/L = 0.086$) in a solution of polymers with length $L_n$ as a function of $L_n/L$. To a very good approximation we find a linear dependence of $\tau_{eq}$ on $L_n$, which excludes a standard constraint-release mechanism based on curvilinear motion of the surrounding chains.
\label{fig:ee_regimes}}
\end{figure}
We find three distinct regimes (Fig.~\ref{fig:ee_regimes}): Initially, the MSD end-to-end distance increases as a power law, $\delta R^2 (t) \,{\sim}\, t^{3/4}$; this agrees with previous experimental observations for freely relaxing semiflexible polymers\cite{leGoffFrey2002prl} as well as corresponding theoretical predictions\cite{farge1993dynamic, frey1991dynamics,  granek1997semi, hallat, kroy1997dynamic}. 
It is followed by an intermediate regime where $\delta R^2 (t)$ increases more slowly, indicating that the internal bending modes are constrained by the surrounding polymers.
We refer to the corresponding time scale marking the crossover from free to constrained relaxation as the `entanglement time' $\tau_e$. 
Finally, we observe at some `equilibration time' $\tau_{eq}$ that the MSD end-to-end distance begins to saturate. 
As the measured saturation value is identical to the thermal equilibrium value\cite{leGoffFrey2002prl, wilhelm1996radial} $\delta R^2_{eq}  \, {=} \, \frac{1}{45} L^4/\ell_p^2$ , the bending modes are now fully relaxed and thus no longer constrained by the surrounding polymers.
In other words, the internal bending modes are disentangled.\\

\textbf{Initial entanglement is described by Odijk-Semenov scaling.} According to the scaling theory for semiflexible polymers by Odijk and Semenov\cite{odijk:83, semenov:86}, topological constraints imposed by neighbouring polymers can effectively be described by a tube of diameter $d  \,{\sim}\,  \xi^{6/5} \ell_p^{-1/5}$, which leads to a restriction of the bending fluctuations of the confined polymer for lengths larger than the `entanglement length' $L_e  \,{\sim}\,  (d^2 \ell_p)^{1/3}$ (for an illustration see Supplementary Movie 2). 
This implies that the dynamics of the end-to-end distance will begin to deviate from free relaxation at the corresponding `entanglement time' $\tau_e  \,{\sim}\,  L_e^4 / \ell_p$. 
All these scaling results for the physics of initial entanglement are in full accordance with our Brownian dynamics simulations (Supplementary Note 2):
To avoid ambiguities, we chose to define the tube diameter $d$ as that value of the transverse fluctuations $g_{1,\bot}(t)$ (see Methods) where it starts to deviate from the $t^{3/4}$-behaviour of free polymers (Supplementary Fig.~7). 
This determines both the tube diameter $d$ (Supplementary Fig.~8) and the entanglement time $\tau_e$, and with $\tau_e \,{\sim}\,  L_e^4 / \ell_p$ also the entanglement length $L_e$ (Supplementary Fig.~9), with all numerical results fully supporting Odijk-Semenov scaling. \\

\textbf{Bending fluctuations equilibrate (disentangle) faster than the diffusion time.} The caging effect of the tube on the bending fluctuations is only transient, as the polymers surrounding the tracer polymer themselves move through bending fluctuations, centre-of-mass and rotational diffusion. 
Hence, at some equilibration time $\tau_{eq}$ the tube will effectively begin to `dissolve' (Supplementary Movie 2, and Supplementary Fig.~10).
Surprisingly, we find that this happens well before the polymer diffuses the full length of the tube, $\tau_{eq}  \,{<}\,  \tau_d$, where the `diffusion time' $\tau_d  \,{=}\,  \zeta L^3 / (6 k_B T)$ is defined as the time a rod takes to diffuse its own length $L$. 
A comprehensive analysis of our Brownian dynamics simulations (Supplementary Note 3) shows that the equilibration time $\tau_{eq}$ is independent of the persistence length (Supplementary Fig.~11) and given by $\tau_{eq}  \,{=}\,  2.7 {\times} 10^{-4} \, L^5 / \xi^2$ (Supplementary Fig.~12).\\
  
\textbf{Constraint release is due to correlated motion of filaments and their neighbouring filaments.} What then drives the release of the $N_\times \,{\sim}\, L / \xi$ topological constraints and thereby determines the equilibration time $\tau_{eq}$? 
As one expects that each of these constraints is released independently, the equilibration time should scale as $\tau_{eq}  \,{\sim}\,  N_\times^2 \, \tau_\times$, where $\tau_\times$ denotes the relaxation time of a single topological constraint. 
At first sight our simulations seem to suggest that the constraint release time equals the diffusion time, $\tau_\times \,{=}\, \tau_d$ (Supplementary Fig.~13), but this is clearly inconsistent with $\tau_{eq}  \,{<}\,  \tau_d$. 
To disentangle these puzzling results, we exploited a unique strength of Brownian dynamics simulations, namely that they allow one to adjust the length and friction coefficient of each individual polymer in the solution (Supplementary Note 3).  
First, we investigated the dynamics of a tracer polymer of length $L$ in an entangled solution of polymers with a shorter length $L_n$, and found $\tau_{eq} \,{\sim}\,  N_\times^2 L_n L^2$ (Fig.~\ref{fig:ee_regimes}b, and Supplementary Fig.~14).
This scaling behaviour can not be explained by a classical constraint release mechanism\cite{mcleish:02, viovy1991constraint} based on sliding motion of the shorter filaments only, as the latter would predict  $\tau_{eq}^\text{cr} \,{\sim}\,  N_\times^2 L_n^3$. 
Instead, our numerical results suggest that the interplay between the dynamics of the tracer polymer and its surrounding polymers drives constraint release. 
To test this hypothesis we varied the relative value of the friction coefficient of the tracer polymer, $\zeta$, and the surrounding polymers, $\zeta_n$, (Supplementary Fig.~15). 
We find $\tau_{eq} \,{\sim}\,  N_\times^2 (\zeta L^3)^\alpha (\zeta_n L_n^3)^\beta$ with $\alpha + \beta = 1$ and the numerical data consistent with $\alpha = 2/3$ and $\beta = 1/3$.
These results are not meant to indicate strict scaling laws but rather to affirm that \textit{correlated motion} of the tracer polymer and its neighbouring polymers is responsible for constraint release.

\begin{figure}[t]
\centering
\includegraphics[width=0.8\linewidth]{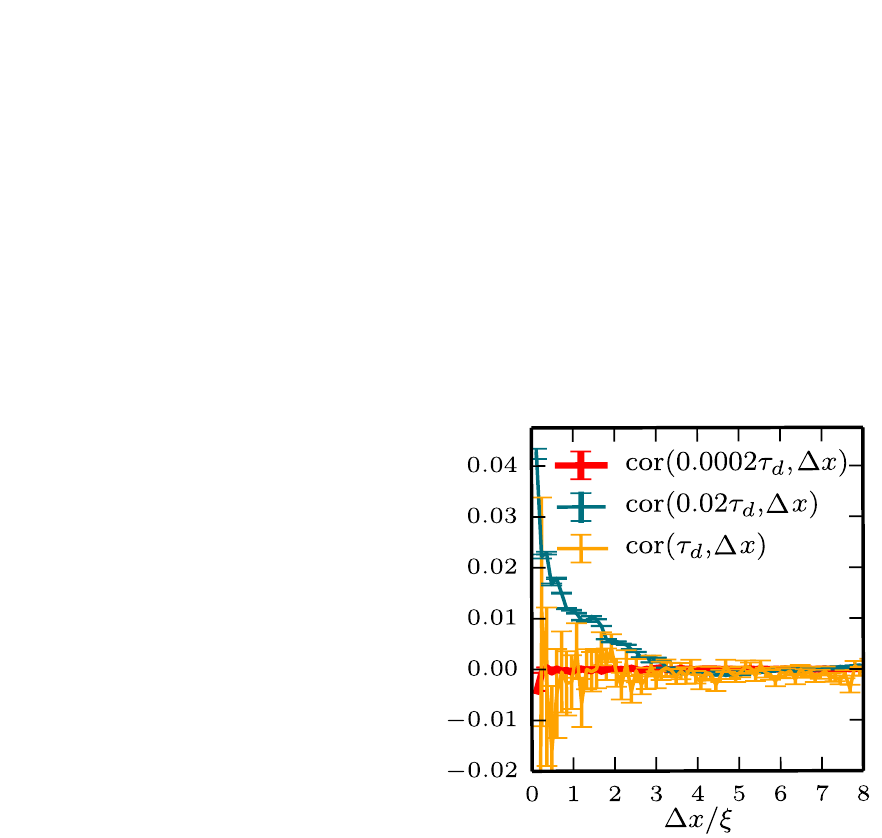}
\caption{\textbf{Spatial correlations of transverse fluctuation.} 
The covariance $\text{cov}(\Delta t, \Delta x)$ of transverse fluctuations $r^k_{\perp, i}$ of beads at position $i$ on neighbouring chains $k$ and $l$ during a time interval $\Delta t$, as a function of the distance $\Delta x \,{=}\,  \left| \mathbf{r}^k_{i}(t\,{+}\,\Delta t) -  \mathbf{r}^l_{i}(t\,{+}\,\Delta t) \right|$, and normalised with respect to $\text{cov}(\Delta t, 0)$: $\text{cor}(\Delta t, \Delta x) \,{:=}\, \text{cov}(\Delta t, \Delta x)/\text{cov}(\Delta t, 0)$. Simulations were performed for an entangled solution with $\xi/L \,{=}\, 0.086$ and $\ell_p /L \,{=}\, 1$, and at different times $t$. As expected, for very small times $t \lesssim \tau_e$ (red curve), there is no correlation. In contrast, for times in the intermediate regime, $\tau_e \,{\lesssim}\, t \,{\lesssim}\, \tau_{eq}$ (here $t \,{=}\, 0.02 \tau_d$), there are clear spatial correlations spanning a distance of about three times the mesh size $\xi$. These correlations vanish for times larger than the equilibration time, e.g.\ for times equal to the diffusion time, $t \,{=}\, \tau_d$.
\label{fig:spatial_correlations}}
\end{figure}

To further substantiate these many-body effects, we measured how the spatial correlations between the polymer's fluctuations perpendicular to its end-to-end distance evolve over time (Fig.~\ref{fig:spatial_correlations}); for a definition of the correlation function see Eq.~\eqref{eq:r_perp}--\eqref{eq:def_orth_cov} in Methods.
Our numerical data show clear spatial correlations in precisely that time window in which tube reorganisation takes place. Moreover, these correlations span several mesh sizes, strongly suggesting that many-body effects are responsible for tube reorganisation and the cocommittant relaxation of internal bending modes.
We have also performed simulations for dilute solutions with $\xi /L \,{=}\, 0.51$ and found no correlations in the transverse displacements of neighbouring chains. 

Taken together, we conclude that constraint release (disentanglement) of internal modes is driven by the correlated motion of a given polymer and its surrounding polymers, and leads to an equilibration of internal bending modes on a time scale $\tau_{eq}$ smaller than the diffusion time $\tau_d$. 
As a consequence, for times larger than $\tau_{eq}$ the tube is dissolved insofar as the bending modes are equilibrated. 
However, this does not suffice for mechanical stresses to relax in a solution of semiflexible polymers, as the latter requires that correlations in the polymer's orientation also must vanish.\\

\begin{figure}[t]
\centering
\includegraphics[width=\linewidth]{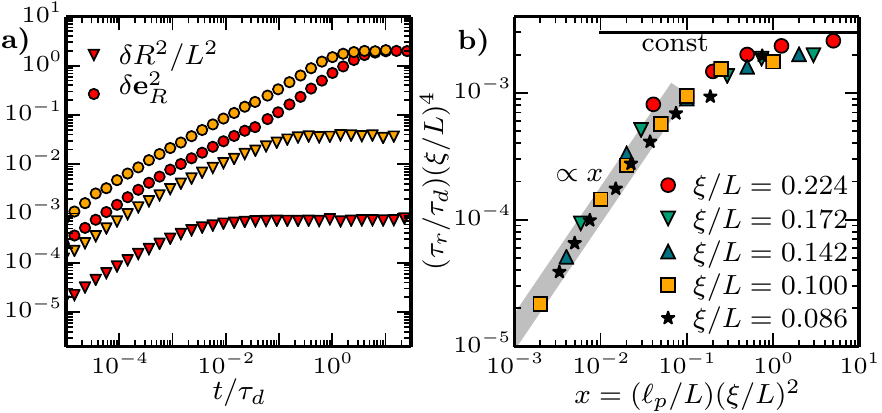}
\caption{\textbf{Terminal relaxation.} 
{\bf a}, Time evolution of the mean-square end-to-end distance $\delta R^2 (t)$ (triangles), and the mean-square changes in the orientation $\delta e_R^2 (t)$ (circles) of a tracer polymer for entangled solutions with $\xi/L \,{=}\, 0.15$, $\ell_p /L \,{=}\, 5$ (red), and $\xi/L \,{=}\, 0.086$, $\ell_p /L \,{=}\, 0.44$ (yellow). Note that terminal relaxation sets in long after relaxation of internal bending modes.
{\bf b}, Scaling plot of the terminal relaxation time $\tau_r / \tau_\text{Doi}$ with $\tau_\text{Doi} \,{=}\, \tau_d (L/\xi^4)$ as a function of $\ell_p/\ell_p^*$ with $\ell_p^* \,{=}\, L^3/\xi^2$, as determined from a fit of the asymptotic relaxation of $\delta e_R^2 (t)$  to the functional form $2{-}2 \exp(-t/\tau_r)$. The numerical data collapse to a master curve with two distinct regimes: For small values of $x \,{=}\, \ell_p/\ell_p^*$, the terminal relaxation time $\tau_r$ increases linearly with $x$, whereas for $ x \,{\gtrsim}\, 1$ it becomes independent of $x$, indicating that one has reached the rigid rod limit where Doi's scaling results are valid\cite{doi1978rods}.
\label{fig:terminal_relaxation}}
\end{figure}

\textbf{Terminal relaxation.} To study this terminal relaxation regime we measured the auto-correlation function for the polymer orientation, $\delta {e}_R^2 (t) \,{:=}\,  \langle \left[ {\bf e} (t) {-} {\bf e} (0) \right]^2 \rangle$, where ${\bf e}(t) \,{=}\,  {\bf R}(t) / R(t)$ denotes the instantaneous, normalised end-to-end vector of the polymer. 
Our simulations show that \textit{terminal relaxation} happens much later than the relaxation of the internal modes (Fig.~\ref{fig:terminal_relaxation}a).
To determine the terminal relaxation time $\tau_r$ we performed a least-squares fit of $\delta {e}_R^2 (t)$ close to saturation to $2{-}2 \exp(-t/\tau_r)$ (Supplementary Fig.~16), as expected for rotational diffusion\cite{granek1997semi}. 
For stiff chains, where the maximal mean-square amplitude of the bending modes $\delta {\bf r}_\perp^2 \,{\sim}\,  L^3 / \ell_p$ is less than the squared mesh size $\xi^2$, we recover Doi's\cite{doi1978rods} result, $\tau_\text{Doi} \,{\sim}\,  \tau_d (L/\xi)^4$ (Fig.~\ref{fig:terminal_relaxation}b).
Hence, in this asymptotic limit, terminal relaxation is facilitated by Doi's \textit{`reptation -- tube rotation'} mechanism\cite{doi1978rods}: after diffusing its own length $L$ within a tube of diameter $d \,{\sim}\,  \xi^2 / L$ in a time of the order of the diffusion time $\tau_d$, the rod becomes confined to a new tube, which is tilted relative to the previous one by an angle $\delta \theta \,{\sim}\,  d/L$. 
The validity of Doi's effective reduction to a single-chain problem in the stiff limit has also been shown recently in computer simulations of needle-like rigid rods\cite{hofling2008entangled, Leitmann_etal:2016}.
In contrast, for $\delta {\bf r}_\perp^2 \,{>}\,  \xi^2$, where tube confinement of bending modes is significant, another relaxation mechanism sets in, which allows terminal relaxation to take place orders of magnitude faster. 
We find $\tau_r \,{\sim}\,  \ell_p L^4 / \xi^2$ (Fig.~\ref{fig:terminal_relaxation}b, and Supplementary Fig.~17), which is clearly distinct from Odijk's\cite{odijk:83} result $\tau_\text{Odijk} \,{\sim}\,  \ell_p L^2$ . 
Moreover, all our simulation data collapse on a universal scaling curve $\tau_r \,{=}\, \tau_\text{Doi} \, \hat \tau (\ell_p / \ell_p^*)$ (Fig.~\ref{fig:terminal_relaxation}b), confirming that there is a crossover from Doi's rigid rod scaling regime\cite{doi1978rods} to a qualitatively different \textit{disentanglement regime} by reducing the persistence length below some threshold value, $\ell_p \,{<}\,  \ell_p^* \,{:=}\,  L^3 / \xi^2$. This constitutes an intermediate asymptotic scaling regime extending over at least two decades ($10^{-3} \,{<}\, x \,{<}\, 1$) in the scaling variable $x = \ell_p / \ell_p^* = \ell_p \xi^2 / L^3$. It is precisely this regime which is most relevant for entangled solutions of cytoskeletal biopolymers\cite{hinner:98} as well as carbon nanotubes\cite{fakhri:10}.

We can explain the observed terminal relaxation building on our results for the equilibration of internal modes by the following physical picture: 
After an initial phase of confinement to an Odijk-Semenov tube, the correlated motion of each polymer and its surrounding polymers leads to a reorganisation of the topological constraints, and thereby to an equilibration of internal bending modes, within a time $\tau_{eq} \,{\sim}\,  L^5/\xi^2$.
As a consequence, there is a mean-square angular rotation of the polymer due to bending fluctuations\cite{wilhelm1996radial, broedersz:14}, $\langle \delta \theta^2 \rangle \,{\sim}\,  L/\ell_p$, and the ensuing rotational diffusion constant scales as $D_r \,{\sim}\,   \langle \delta \theta^2 \rangle / \tau_{eq} \,{\sim}\,  \xi^2 / (L^4 \ell_p)$. 
This is identical to the scaling of the terminal relaxation time as we find it in our simulations, $\tau_r \,{\sim}\,  \ell_p L^4 / \xi^2$ (Fig.~\ref{fig:terminal_relaxation}b).
Hence, terminal relaxation in solutions of entangled semiflexible polymers is not primarily due to diffusive motion along the tube's backbone (reptation), but rather due to \textit{disentanglement} of internal bending modes mediated by the release of topological constraints and the ensuing rotational diffusion of the polymer's orientation, induced by bending modes ($\langle \delta \theta^2 \rangle \,{\sim}\,  L/\ell_p$). 
The constraint release for the internal modes is facilitated by the correlated dynamics of each polymer and its surrounding polymers. \\

\textbf{Comparison of entangled solutions with frozen environments.} 
In order to better understand the origin of these differences and the role of many-body effects in the dynamics of entangled solutions, we also studied the dynamics of a reference system where a tracer polymer moves in a frozen environment, similar as in recent experiments studying the Brownian motion of carbon nanotubes in a fixed agarose network\cite{fakhri:10}. 
In good quantitative agreement between our simulations in the frozen environment and the experimental results\cite{fakhri:10}, the rotational relaxation time exhibits a crossover from Doi scaling\cite{doi:78} to Odijk scaling\cite{odijk:83} with increasing polymer length $L$ (Fig.~\ref{fig:frozen}): 
\begin{figure}[t]
\centering
\includegraphics[width=\linewidth]{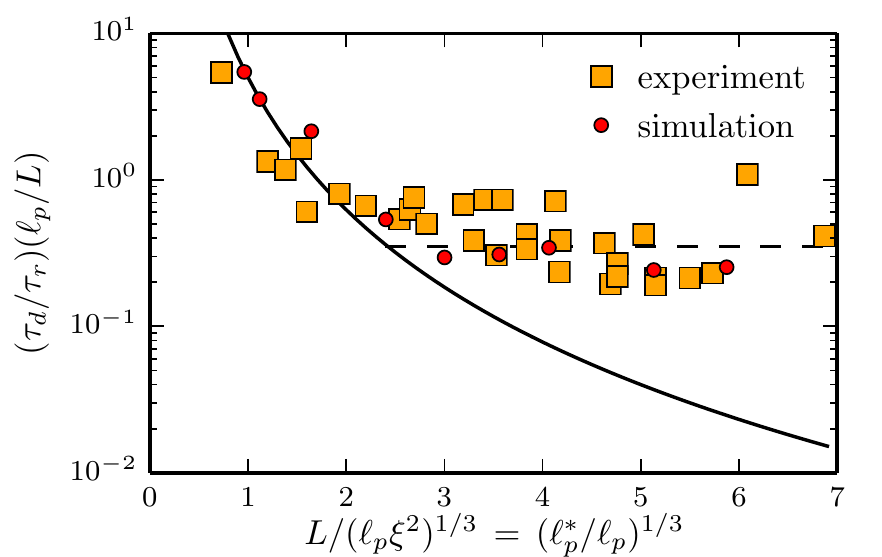}
\caption{\textbf{Scaling of the rotational relaxation time for a tracer polymer in an array of frozen polymer chains.}
The numerical results for the rotational diffusion constant $D_r \,{\sim}\, \tau_r^{-1}$ in systems of different density and flexibility, rescaled as $(\tau_d/\tau_r) (\ell_p/L)$, are plotted as a function of rescaled polymer length $L/(\ell_p \xi^2)^{1/3} \,{=}\, (\ell_p^*/\ell_p)^{1/3}$. 
As in the experimental system, where carbon nanotubes diffuse in a porous agarose network (squares, data extracted from Ref.\cite{fakhri:10}), our simulations (solid circles) show that with increasing filament length $L$ there is a crossover from Doi's scaling result for rigid rods, $\tau_\text{Doi} \,{\sim}\, L^7/\xi^4$, (solid line) to Odijk's scaling result for semiflexible polymers, $\tau_\text{Odijk} \,{\sim}\,  \tau_\text{Doi} \, \ell_p / L \,{\sim}\, \ell_p  L^2$. To fit our definition, which uses the saturation time $\tau_r$ instead of the diffusivity, the experimental data has been rescaled by the necessary factor of $0.5$. Our simulations agree quantitatively well with the experimental data\cite{fakhri:10}.
\label{fig:frozen}}
\end{figure} 
For rather stiff polymers with $ L/(\xi^2 \ell_p)^{1/3}  \,{\lesssim}\, 2$, the relaxation time agrees with Doi's prediction, $\tau_r \,{\sim}\, L^7 / \xi^4$, while for semiflexible chains with $ L/(\xi^2 \ell_p)^{1/3}  \,{\gtrsim}\, 2$, one finds Odijk's prediction, $\tau_r \,{\sim}\,  \ell_p L^2$. 
Moreover, our Brownian dynamic simulations in a frozen environment are also consistent with previous numerical results of rigid rods in a random array of fixed obstacles where $\tau_r$ is reported to follow Doi's prediction\cite{munk:09, hofling2008enhanced, hofling2008entangled, nam2010reptation}.

\begin{figure}[!t]
\centering
\includegraphics[width=0.835\linewidth]{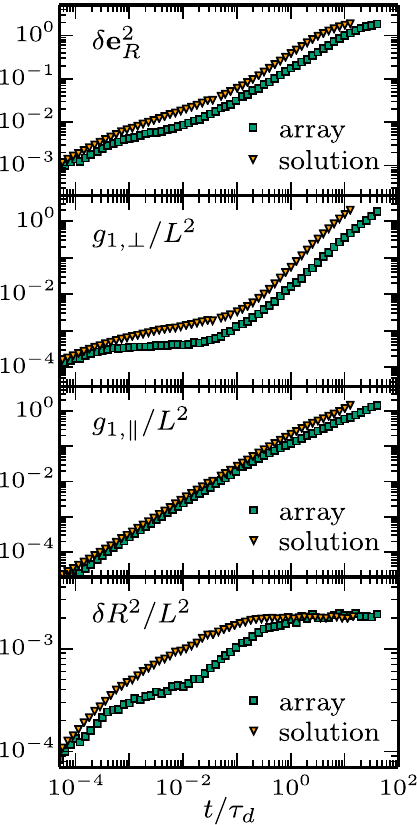}
\caption{\textbf{Comparison between dynamics in a fixed array and an entangled solution}. Key quantities characterising the dynamics of a tracer polymer in a fixed array (green squares) or an entangled solution (yellow triangles) for a system with $\xi/L \,{=}\, 0.086$ and $\ell_p /L \,{=}\, 3$. To emulate a fixed array we used an equilibrated configuration of the entangled solution and froze the configuration of all but one tracer polymer, which is reminiscent of the experimental situation in Ref.\cite{fakhri:10}. In a fixed array, relaxation of a tracer polymer's orientation ($\delta e_R^2$) is much slower and follows the scaling predictions of Odijk and Doi. Moreover, the MSD of the polymer's fluctuations transverse to the end-to-end vector ($g_{1,\bot}$) exhibits a clearly visible plateau in a fixed array, but continues to relax in an entangled solution. At asymptotically large times, the transverse MSD and likewise the longitudinal MSD ($g_{1,\parallel}$) is slowed down in a fixed array. Finally, the relaxation of internal bending modes ($\delta R^2$) shows a qualitatively different behaviour: While there is an intermediate regime where $\delta R^2 (t)$ appears to saturate, which reflects caging in a tube, this is not the case for an entangled solution. There internal modes continuously relax, indicating ongoing tube reorganisation with concomitant relaxation of topological constraints. 
\label{fig:fixed_vs_solution}}
\end{figure}
Comparing the dynamics in a fixed environment with the dynamics in an entangled solution, we find that \textit{all} key observables show qualitatively different behaviour 
(Fig.~\ref{fig:fixed_vs_solution}): 
The relaxation of the internal modes and the orientational correlations is retarded in a frozen environment, and the equilibration time $\tilde \tau_{eq}$, as well as the terminal relaxation time $\tilde \tau_r$, now agree well with the time it takes to diffuse the tube length, $\tilde \tau_{eq} \,{=}\,  \tau_d$, and Odijk's scaling result, $\tau_\text{Odijk} \,{\sim}\,  \ell_p L^2$, respectively. 
While for an entangled solution the correlation function for the transverse displacement ($g_{1,\perp}$) of the center of mass exhibits a slanted plateau, it shows a clear flat plateau in a frozen environment, indicating that the dynamics of the tracer polymer is confined to a persistent tube in the time window between the entanglement time $\tau_e$ and the tube diffusion time $\tau_d$. 
These predictions for the various key observables may be tested using microrheology experiments.
Taken together, the comparison between the dynamics of entangled solutions and fixed arrays reaffirm that tube reorganisation, and the ensuing relaxation of topological constraints, must be based on the combined effect of the dynamics of a given polymer and its surrounding polymers.

\section*{Discussion}

The results presented here challenge our current understanding of the dynamics of entangled polymer solutions. 
While both asymptotic regimes, flexible polymers and rigid rods, are well described within the framework of classical reptation theory, our current data show that semiflexible polymer dynamics clearly exhibits strong many-body correlation effects. 
These correlation effects lead to a fast equilibration ('disentanglement') of the internal bending modes significantly before a polymer had time to diffuse the full length of its confining tube.  
As a consequence, terminal relaxation is facilitated by the correlated release of topological constraints and the ensuing rotational diffusion of the polymer orientation, and not by curvilinear motion as in classical reptation theory. 
The processes responsible for the restructuring and renewal of the tube appear to show similarities with dynamic correlations found in the glassy behaviour of dense colloidal systems\cite{berthier2011theoretical}. 
However, it remains a challenge to identify the essence of the many-body physics responsible for tube renewal leading to the relaxation of bending modes.

To test the proposed disentanglement mechanism underlying terminal relaxation experimentally one actually does not need to measure the terminal relaxation time explicitly. 
It suffices to measure the equilibration time of the internal modes and validate the predicted scaling of the equilibration time.
This should be feasible well within the time window accessible to experiments on entangled solutions of cytoskeletal filaments, carbon nanotubes, or short DNA filaments.

\begin{acknowledgements}
This project was supported by the German Excellence Initiative via the program ``NanoSystems Initiative Munich'' (NIM).
\end{acknowledgements}

\label{methods}

\section*{Methods}

\textbf{Brownian dynamics simulation.} 
We implemented a basic bead-spring algorithm\cite{fixman:78, grassia:95} to simulate the Brownian dynamics of polymer chains in an entangled solution, using standard interactions as discussed previously\cite{kremer1990dynamics, kremer1992simulations, paul1991dynamics, nam2010reptation}, and explained in detail below. 
Our simulations comprise $M$ polymer chains with $164 \,{\leq}\, M \,{\leq}\, 1106$ in a cubic simulation volume with edge length $X \,{=}\, 1.35\, L$, and periodic boundary conditions. Each polymer is represented by a linear sequence of $N$ beads connected by springs. 
For low densities we used $N \,{=}\, 45$, while for densities above $\xi/L \,{\leq}\, 0.1$ we used a finer discretisation with $N \,{=}\, 60$; this ensures that the polymers are thin enough such that the density of the entangled solution is sufficiently below the threshold to the nematic phase\cite{dijkstra1995simulation, khokhlov1981liquid, onsager:49}.

In all our simulations we chose units such that $k_BT \,{=}\, 1$, and the friction per unit length is $\zeta \,{=}\, 1$. 
Also, we used the contour length of the polymers as our basic unit of length. 
For later reference, an actin filament with a contour length of $L \,{=}\, 10 \, \mu \mathrm{m}$ and a diameter of $5 \, \mathrm {nm}$ in a solution with a viscosity of $\eta \,{=}\, 0.1 \, \mathrm {Pa \, s}$ at a temperature of $20^\circ C$ has a disengagement time of $\tau_d \,{\approx}\, 1.5 \times 10^4 \, \mathrm{s}$.

The position of bead $i$ on polymer chain $k$ at time $t$ is denoted by $\mathbf{r}_{i}^{k}(t)$ with $1 \,{\leq}\, i \,{\leq}\, N$ and $1 \,{\leq}\, k \,{\leq}\, M$. 
Each bead $i$ is connected to its neighbours by FENE springs\cite{grest1986} with the interaction potential
\begin{equation}
\mathcal U^k_\text{FENE} 
= -B \sum_{1<i<N-1} 
\ln 
\left[ 
1-\left( \frac{a^k_i-a_0}{a_\text{max}} \right)^2   
\right] \, ,
\end{equation}
where $a^k_i \,{=}\, |\mathbf{r}_{i+1}^{k} {-}\mathbf{r}_{i}^{k} |$ denotes the distance between two beads (bond length), with $a_0 \,{=}\, L/N$ the equilibrium bond length, and the maximum distance between beads set to $a_\text{max} \,{=}\, a_0/4$; for distances $|a^k_i| \,{>}\, a_0 \,{+}\, a_\text{max}$ the FENE potential is $\mathcal U^k_\text{FENE} \,{=}\, \infty$. 
We chose the spring constant $B \,{\approx}\, 1000 \, k_BT$. With this choice the deviation in bond length from the equilibrium value was always below $0.05 \, a_0$ in our simulations. 
The bending stiffness of the polymers is described by a standard worm-like chain model\cite{Kratky:1949, Saito:1967} with the bending energy given by 
\begin{equation}
\mathcal U^k_{\mathrm{WLC}} 
= \frac{\ell_p k_BT}{a_0} 
\sum_{i=1}^{N-2} 
\left( 1 - \mathbf{t}^k_i \cdot \mathbf{t}^k_{i+1} \right) 
\, ,
\end{equation}
where $\mathbf{t}^k_i \,{=}\, ({\mathbf{r}_{i+1}^{k} -\mathbf{r}_{i}^{k}})/{|\mathbf{r}_{i+1}^{k} -\mathbf{r}_{i}^{k}|}$ is a normalised bond vector (tangent vector). 
For the mutual (steric) interaction between the beads we used the Weeks-Chandler-Anderson (WCA) potential\cite{weeks1971role}, which for bead $i$ on chain $k$ reads for $r_{ij} \,{\leq}\, \sigma$
\begin{equation}
\mathcal U^{i,k}_{\mathrm{WCA}} 
=  A  \sum\limits_{l,j}  
\left[ 
  \left(\frac{\sigma}{r^{kl}_{ij}} \right)^{12} - 
2 \left(\frac{\sigma}{r^{kl}_{ij}} \right)^{6} + 1
\right] \, ,
\end{equation}
% $\sum\limits_{\substack{1<l<M \\ 1<j<N \\ j \neq i \ \mathrm{for}\ k=l} }$
where the sum extends over all other beads in the simulation box, i.e.\ over all polymers $l$ and beads $j$ except bead $i$ of chain $k$, and $r^{kl}_{ij}  \,{=}\,  |\mathbf{r}_{i}^{k} \,{-}\, \mathbf{r}_{j}^{l}|$ denotes the distance between a pair of beads $(i,j)$ on chains $k$ and $l$. 
For distances $r_{ij} \,{>}\, \sigma$, the potential vanishes: $\mathcal U^{i,k}_{\mathrm{WCA}} \,{=}\, 0$. 
We chose $\sigma \,{=}\, 0.9 \, a_0$ for the range of the interaction.
This choice prevents chain crossings and at the same time spurious oscillations of neighbouring beads in a chain due to their WCA interaction. 
In order to avoid crossing of chains or the overlapping of different beads we used a strong potential by setting the parameter $A  \,{=}\,  20 \, k_B T$ (see Supplementary Note 1). 
Finally, we implemented a cell-linked list algorithm to evaluate the occurring collisions which presorts the beads according to their positions before testing for polymer collisions. 
This led to a significant decrease in the runtime of our simulations. 

The Langevin equation for the entangled polymer solution reads 
\begin{eqnarray}
\label{eq:eom}
 \zeta \, \frac{\partial \mathbf{r}^k_{i}(t)}{\partial t} = - \frac{\partial \, \mathcal{U}^{i,k}_{\mathrm{total}} ( \{\mathbf{r}^k_{i} \})}{\partial \mathbf{r}^k_{i}} + \boldsymbol{\eta}^k_i(t) \, , 
\label{eq:langevin}
\end{eqnarray}
where $\zeta$ denotes the friction coefficient of a single bead, $\boldsymbol{\eta}^k_i(t)$ is Gaussian white noise with mean zero and co-variances given by 
$
\langle \boldsymbol{\eta}^k_i(t) \, \boldsymbol{\eta}^l_j(t') \rangle 
= 6   \, \zeta\, \delta_{ij} \delta_{kl} \, k_B T \, \delta(t-t')$.
The total potential acting on bead $i$ of polymer $k$ reads: $\mathcal{U}^{i,k}_{\mathrm{total}} \,{:=}\, U^{i,k}_{\mathrm{WCA}} \,{+}\,\mathcal U^k_\text{FENE} \,{+}\, \mathcal U^k_{\mathrm{WLC}}$.

We used uniformly distributed random numbers generated by a maximally equidistributed combined Tausworthe generator\cite{galassi2011gsl} for the noise. These kinds of random number generators have been shown to amount to the same behaviour in the dynamics of polymers on time scales significantly above one time step as Gaussian white noise within the statistical errors while being significantly faster\cite{grassia:95, kremer1992simulations, ramanathan:07}. To calculate the time evolution, the Langevin equation, Eq.~\eqref{eq:langevin}, is integrated via a semi-implicit Euler algorithm\cite{euler} with a time step of (in our units) $1 \,{\times}\, 10^{-4}$ for systems with $N \,{=}\, 45$ and $2 \,{\times}\, 10^{-5}$ for $N \,{=}\, 60$, respectively.

We performed extensive tests to ensure the reliability of our Brownian dynamics simulations (Supplementary Note 1). We find good agreement of both the tangent-tangent correlations (Supplementary Fig.~1) and the mean-square end-to-end distance (Supplementary Fig.~2) for freely relaxing polymers with known analytical results\cite{Kratky:1949, farge1993dynamic, frey1991dynamics,  granek1997semi, hallat, kroy1997dynamic}. Moreover, we have tested that our simulation algorithm does not show spurious chain crossings (Supplementary Fig.~3). Finally, we have tested for finite size effects (Supplementary Fig.~4), and that our results for the mean-square displacement of the center monomer (Supplementary Fig.~5) and the mean-square changes of the end-to-end-vector are largely independent of the finite filament thickness (Supplementary Fig.~6). 

\textbf{Quantities of interest.}
To characterise the dynamics of individual chains within the entangled polymer solution we studied the following quantities of interest where the averages $\langle \ldots \rangle$ indicate averages over all $M$ polymers in the simulation box, and over three independent realisations. We chose to characterise the dynamics of the internal bending modes of a tracer polymer in terms of its tangent-tangent correlation function,
\begin{equation}
C_{ij}
:= \langle \mathbf{t}_i \cdot \mathbf{t}_{j} \rangle \, ,
\end{equation}
and the mean-square displacements (MSD) of the end-to-end distance $R^k = |{\bf R}^k|$,
\begin{equation}
\delta R^2 (t) 
:= \langle \left[ R^k(t) -R^k(0)  \right]^2 \rangle \, .
\end{equation}
In order to characterise the center-of-mass motion of a tracer filament we used the mean-square displacement of the center monomer parallel and perpendicular to the orientation of the end-to-end vector ${\bf e}^k \,{=}\,  {\bf R}^k / R^k$, respectively:
\begin{eqnarray}
g_{1,\parallel}(t)  
& := &\left\langle \left[  \left(\mathbf{r}^k_{N/2}(t)  -\mathbf{r}^k_{N/2}(0)  \right)    \cdot \mathbf{e}^k (0)  \right]^2 \right\rangle \, ,
\\
g_{1,\bot}(t)  
& := & \left\langle \left[ \mathbf{r}^k_{N/2}(t)  -\mathbf{r}^k_{N/2}(0)  \right]^2 \right\rangle - g_{1,\parallel} (t) \, .
\end{eqnarray}
Finally, a quantity which allows to measure the terminal relaxation of stresses in the solution is given by the \textit{mean-square changes of the direction of the end-to-end vector}
\begin{equation}
\delta {e}_R^2 (t) \,{:=}\,  \langle \left[ {\bf e}^k (t) {-} {\bf e}^k (0) \right]^2 \rangle \, .
\label{eq:MSD_rot}
\end{equation}

\textbf{Definition of the covariance for transverse displacements.}
To quantify the many-body effects in entangled polymer solutions, we were looking for correlations in the dynamics of neighbouring polymers. Since the tube is mainly constraining the fluctuations of a polymer transverse to its end-to-end vector, and the tube itself is due to the presence of neighbouring chains, we investigated correlations in these transverse fluctuations. We considered the magnitude of the displacement of bead $i$ on polymer $k$ during a time interval $\Delta t$, perpendicular to polymer's end-to-end vector  $\mathbf{e}^k (t \,{+}\,\Delta t)$
\begin{equation}
r^k_{\perp, i} := 
\left| 
\mathcal{P}_\perp^k (t+\Delta t) \cdot \left[ \mathbf{r}^k_{i}(t)-
       \mathbf{r}^k_{i}(t+\Delta t) 
\right] 
\right| \, ,
\label{eq:r_perp}
\end{equation}
where 
\begin{equation}
\mathcal{P} _\perp^k (t) = 
1-\mathbf{e}^k(t) \otimes \mathbf{e}^k (t)
\end{equation}
is a projection operator onto the end-to-end vector. Similar to the work of Doliwa and Heuer\cite{doliwa1998cage} on the cage effect in colloidal systems, we asked for correlations in the transverse displacements $r^k_{\perp, i}$ of neighbouring polymers, and define their covariance as 
%\cite{berthier2011theoretical, doliwa1998cage} 
\begin{equation}
\text{cov}(\Delta t, \Delta x)  := 
\left\langle  
r^k_{\perp, i}  
r^l_{\perp, i}
\right\rangle   
- \langle r^k_{\perp, i} \rangle 
  \langle r^l_{\perp, i} \rangle \, .
\label{eq:def_orth_cov}
\end{equation}
The average in Eq.~\eqref{eq:def_orth_cov} is taken for a given chain $k$ with all other chains $k \,{\neq}\, l$ at a given time $\Delta t$ and a distance $\Delta x \,{=}\,  \left| \mathbf{r}^k_{i}(t\,{+}\,\Delta t) -  \mathbf{r}^l_{i}(t\,{+}\,\Delta t) \right|$, and we have also performed a moving time window average over a window of size $15 \, \tau_d$ discretised in subintervals of size $\Delta t$. Moreover, for simulations with $\Delta t \,{<}\, 0.2 \tau_d$ and $\Delta t \,{=}\, \tau_d$, we averaged over $4$ or $12$ independent realisations, respectively. For specificity, we used the beads at position $i \,{=}\, N/5$ and the equivalent beads at $i \,{=}\, 4N/5$.

\end{document}